\documentclass[twocolumn,showpacs,preprintnumbers,amsmath,amssymb]{revtex4}
\font\mybb=msbm10 at 10pt
\def\bb#1{\hbox{\mybb#1}}
\begin{document}

\title{BCFW-type recurrent relations
for tree superamplitudes of D=11 supergravity}

\author{Igor  Bandos
$^{\dagger\ddagger}$
}
\address{$^{\dagger}$Department of
Theoretical Physics, University of the Basque Country UPV/EHU,
P.O. Box 644, 48080 Bilbao, Spain
 \\ $^{\ddagger}$
IKERBASQUE, Basque Foundation for Science, 48011, Bilbao, Spain }

\date{V1: April 30, 2016. V3. Sept. 27, 2016. Phys.Rev.Lett. {\bf 118}, 031601 (2017) January 20, 2017}

\begin{abstract}

We propose the on-shell superfield description for tree amplitudes  of D=11 supergravity and the BCFW (Britto-Cachazo-Feng-Witten)-type recurrent relations for these superamplitudes.

\end{abstract}


\maketitle


Recent years we are witnesses of a great progress in calculations of multiloop amplitudes (see e.g.
\cite{Bern:2011qn,Drummond:2008vq,Benincasa:2013faa,Kallosh:2014hga} and refs. therein) an important  part of which is related to the applications and development of the Britto-Cachazo-Feng-Witten (BCFW) approach \cite{Britto:2005fq}. This first allowed to obtain Britto-Cachazo-Feng (BCF) recursion relations for tree amplitudes in D=4 Yang Mills and  ${\cal N}=4$ supersymmetric Yang-Mills  (SYM) theory   \cite{Britto:2004ap,Luo:2005rx,Bianchi:2008pu}  and then was developed for the case of superamplitudes of ${\cal N}=4$ SYM   \cite{Brandhuber:2008pf,ArkaniHamed:2008gz}, loop (super)amplitudes and ${\cal N}=8$ supergravity  \cite{Brandhuber:2008pf,ArkaniHamed:2008gz,Heslop:2016plj,Herrmann:2016qea}  (see \cite{Heslop:2016plj,Herrmann:2016qea} for more references). To lighten the text, below we will mainly omit 'super' in superamplitudes, calling them amplitudes.

This approach was generalized for the tree amplitudes of  D=10 SYM model in \cite{CaronHuot:2010rj}, but then  mainly used in the context of type IIB supergravity
\cite{Boels:2012ie,Boels:2012zr,Wang:2015jna,Wang:2015aua} where the presence of complex structure allowed to lighten the `Clifford superfield' description of amplitudes  in \cite{CaronHuot:2010rj}.
The observation that the constrained bosonic spinor helicity variables used in \cite{CaronHuot:2010rj} can be identified with spinor moving frame variables of \cite{BZ-str,BZ-M2,Bandos:1996ju} (or equivalently, with Lorentz harmonics of \cite{Galperin:1991gk,Delduc:1991ir}) \footnote{This observation was also made independently in \cite{Uvarov:2015rxa} and used their to develop the spinor helicity formalism for D=5 massless fields. The D=6 spinor helicity formalism was developed in \cite{Cheung:2009dc}.}  allowed us to simplify it('s ${\cal N}=1$ version) \cite{Bandos:In-prep} and also to generalize it to the case of $D=11$ supergravity\footnote{See \cite{Green:1997as,Anguelova:2004pg,Green:2016tfs} and refs. therein   for other approaches to amplitudes of D=11 supergravity. }. The results of this 11D generalization of the on-shell  superfield description of tree amplitudes and of the BCFW recurrent relations for these will be reported in this letter.

The BCFW  recursion relations \cite{Britto:2005fq} are written for $n$-particle tree  amplitudes ${\cal A}^{(n)}(p_{(1)},\varepsilon_{(1)}; ...,p_{(n)},\varepsilon_{(n)})$ in spinor helicity formalism, in which the information on the (light-like) momentum $p_{\mu (i)}$  and on helicity of the $i$-th external particle are encoded in the bosonic spinor
$\lambda^{A}_{(i)}= (\bar{\lambda}{}^{{\dot{A}}}_{(i)})^*$. The light-like momentum is defined  by  Cartan-Penrose representation   (see \cite{Penrose:1972ia} and refs. therein)
\begin{eqnarray}\label{p=ll=4D}
 p_{\mu (i)} \sigma^\mu_{{A\dot{A}}}=  2\lambda_{A(i)}  \bar{\lambda}_{{\dot{A}(i)}}  \qquad \Leftrightarrow && \; p_{\mu(i)}= \lambda_{(i)} \sigma_\mu \bar{\lambda}_{(i)},  \qquad
\end{eqnarray}
where $\sigma^\mu_{_{A\dot{A}}}$ are relativistic Pauli matrices, $ A=1,2$ and $\dot{A}=1,2$ are Weyl spinor indices and $ \mu=0,...,3$.

All  n-particle amplitudes for the fields of the ${\cal N}$=4 SYM  can be described by a superfield amplitude (superamplitude) \cite{Brandhuber:2008pf,ArkaniHamed:2008gz}  ${\cal A}^{(n)}(\lambda_{(1)}, \bar{\lambda}_{(1)}, \eta_{(1)}; ...;\lambda_{(n)}, \bar{\lambda}_{(n)},  \eta_{(n)})$ depending, besides
 $\lambda^{A}_{(i)}$ and $\bar{\lambda}{}^{^{\dot{A}}}_{(i)}$, on the set of  $n$  complex fermionic coordinates $\eta_{(i)}^q=(\bar{\eta}_{q(i)})^*$ (first introduced in  \cite{Ferber:1977qx}), $\; \eta_{(i)}^q\eta_{(j)}^p=-\eta_{(j)}^p\eta_{(i)}^q$, $\; \bar{\eta}_{q(i)}\eta_{(j)}^p=-\eta_{(j)}^p\bar{\eta}_{q(i)}$, carrying the index $q=1,...,4$ of the fundamental representation of $SU(4)$. These  superfield amplitudes are  multiparticle counterparts of the so-called on-shell superfield
 \begin{eqnarray}\label{Phi=3,4}
&  \Phi (\lambda ,\bar{\lambda}, \eta^q) = f^{(-)}(\lambda ,\bar{\lambda}) +  \eta^q \chi_q + \frac 1 2 \eta^q \eta^p s_{pq}+\nonumber \qquad \\ &
 + \frac 1 {3!} \eta^q \eta^p\eta^r \epsilon_{rpqs}\bar{\chi}^s + \frac 1 {4!} \eta^q \eta^p\eta^r\eta^s  \epsilon_{rpqs}  f^{(+)}
  \quad \end{eqnarray}
 describing all the states  of the linearized SYM {\it provided} it obeys the so-called helicity constraint \cite{Penrose:1972ia,Ferber:1977qx},
 \begin{eqnarray}\label{UPhi=1}
 & \hat{h}  \Phi (\lambda ,\bar{\lambda}, \eta) =
 \Phi (\lambda ,\bar{\lambda}, \eta) \; , \qquad \\
 \label{U=}
 & 2\hat{h}:=
 -\lambda^A \frac \partial  {\partial \lambda^A }+
 \bar{\lambda}{}^{{\dot{A}}} \frac \partial  {\partial \bar{\lambda}{}^{{\dot{A}}}} +
 \eta^q \frac \partial  {\partial  \eta^q} \; .  \qquad \end{eqnarray}
 The $n$-particle on-shell superfield amplitudes of 4D ${\cal N}$=4 SYM,  $  {\cal A}^{(n)}(\lambda_{(1)}, \bar{\lambda}_{(1)}, \eta_{(1)}; ...;\lambda_{(n)}, \bar{\lambda}_{(n)},  \eta_{(n)})\equiv {\cal A}^{(n)}(...; \lambda_i, \eta_i;...)$, should obey the set of $n$ helicity constraints,
 \begin{eqnarray}\label{UcAn=cAn}
\hat{h}_{(i)} {\cal A}^{(n)} (...; \lambda_i, \bar{\lambda}_{i}, \eta_i;...) =  {\cal A}^{(n)}(...; \lambda_i, \bar{\lambda}_{i}, \eta_i;...) \, ,  \qquad  \end{eqnarray}
with $
2\hat{h}_{(i)}:=
 -\lambda^A_{(i)} \frac \partial  {\partial \lambda^A_{(i)} }+
 \bar{\lambda}{}^{{\dot{A}}}_{(i)} \frac \partial  {\partial \bar{\lambda}{}^{{\dot{A}}}_{(i)}} +
 \eta^q_{(i)} \frac \partial  {\partial  \eta_{(i)}^q}
$.

We refer to \cite{Brandhuber:2008pf,ArkaniHamed:2008gz}  for the superfield generalization of the   original D=4 BCFW recurrent relations \cite{Britto:2005fq}, and pass to the 11D generalization of the spinor helicity formalism.

\bigskip

\centerline{\bf 1. Spinor helicity formalism in D=11.}

Let us denote the D=11 vector indices by $a,b,c=0,1,...,9,10$, spinor indices of SO(1,10) by $\alpha,\beta, \gamma, \delta=1,...,32$ and D=11 Dirac matrices by $\Gamma_{a\alpha}{}^{\beta}$. In our  mostly minus notation,
$ \eta^{ab}=diag (+1,-1,...,-1)$, both $\Gamma_{a\alpha}{}^{\beta}$ and the charge conjugation matrix
  $C^{\alpha\beta}=-C^{\beta\alpha}$ are imaginary. We will also use the real symmetric matrices
  $\Gamma^a_{\alpha\beta}=\Gamma_{\alpha}^{a\gamma} C_{\gamma\beta}= \Gamma^a_{\beta\alpha}$, and $\tilde{\Gamma}{}_{a}^{\alpha\beta}=C^{\alpha\gamma} \Gamma_{\gamma}{}^{a\beta}= \tilde{\Gamma}{}_{a}^{\beta\alpha} $.

The light-like momentum of a massless 11D particle can be expressed by the relations  similar to (\ref{p=ll=4D}),
\begin{eqnarray}\label{k=pv-v-11}
k_a
\Gamma^a_{\alpha\beta}= 2\rho^{\#} v_{\alpha q}^{\; -} v_{\beta q}^{\; -}  \; , \qquad
 \rho^{\#} v^-_{{q}} \tilde{\Gamma}_{a}v^-_{{p}}= k_{a} \delta_{{q}{p}}\; , \qquad
\end{eqnarray}
in terms of 'energy variable' $\rho^{\#} $ and a set of 16 constrained bosonic 32-component spinors $v_{\alpha q}^{\; -}$, $q,p=1,..., 16$, which can be identified with D=11 spinor moving frame variables \cite{Bandos:2006nr,Bandos:2007mi,Bandos:2007wm}  or  Lorentz harmonics \cite{Galperin:1992pz}. Essentially,  the constraints on $v_{\alpha q}^{\; -}$  are given by Eq. (\ref{k=pv-v-11}) supplemented by $v_{\alpha q}^{\; -}C^{\alpha\beta}v_{\beta q}^{\; -}=0$,
and by the requirement that the rank of $32\times 16$ matrix $v_{\alpha q}^{\; -}$ is equal to 16. We refer to  \cite{Bandos:2007mi,Bandos:2007wm} for the complete description and discussion of the constraints and gauge symmetries of the spinor moving frame formalism for 11D massless superparticle and only notice  that, taking all these into account,
the variables  $v_{\alpha q}^{\; -}$ can be considered as  homogeneous coordinates on ${\bb S}^{9}$,
the  celestial sphere of a D=11 observer,
\begin{eqnarray}\label{v-=S9}
 \{ v_{\alpha q}^{\; -}\} = {\bb S}^{9}\; . \qquad
\end{eqnarray}
The sign superindices $^{-}$ and $^{\#}\equiv ^{++}$, carried by $v_{\alpha q}^{\; -}$ and $\rho^{\#}$, characterize their scaling properties with respect to $SO(1,1)$ gauge symmetry of the spinor moving frame (or Lorentz harmonic) approach to massless (super)particle.

One can check that, due to (\ref{k=pv-v-11}) and $v_{q}^{-}Cv_{p}^{-}=0$, the momentum vector $k_a$ is light-like, $\; k_ak^a=0\,$, and moreover that the spinor moving frame variables  $v_{\alpha q}^{\; -}$ obey the massless Dirac equation (in momentum representation)
\begin{eqnarray}\label{Dirac}
 & k_{a} \tilde{\Gamma}{}^{a\, \alpha\beta}v_{\beta  q}{}^- =0 \qquad \Leftrightarrow \qquad k_a\Gamma^a_{\alpha\beta}  v^{-\beta}_{q}=0\;  . \qquad
\end{eqnarray}

The 11D counterpart of the 10D spinor helicity variables of \cite{CaronHuot:2010rj} are $
\lambda_{\alpha q}= \sqrt{\rho^{\#}} v_{\alpha  q}^{\; -}$;
the counterpart of the polarization spinor of the 10D fermionic field in D=11 is given by the same helicity spinor but with risen spinor index,  $\lambda_q^{\alpha}= \sqrt{\rho^{\#}} v_{q}^{-\alpha  }=-i C^{\alpha\beta}\lambda_ {\beta q} $ ($= (\lambda_q^{\alpha})^*$).

One notices that Eqs. (\ref{k=pv-v-11}) can be written as {$\Gamma^a_{\alpha\beta}k_{a}= 2\lambda_{\alpha q}\lambda_{\beta q}$ and $
\lambda_{{q}} \tilde{\Gamma}_{a}\lambda_{{p}}= k_{a} \delta_{{q}{p}}$.
However,  the energy variable $\rho^{\#}$ and its canonically conjugate coordinate  $x^=$ play an important role in our construction below. In particular the D=11 counterpart of the on-shell superfields are defined on superspace
\begin{eqnarray}\label{On-shellSSP}
\Sigma^{(10|16)} : \qquad &&  \{ (x^= ,  v_{\alpha q}^{\; -}; \theta^-_q)\} \; ,  \qquad
\end{eqnarray}
with bosonic sector ${\bb R}\otimes {\bb S}^{9}$ (see (\ref{v-=S9})) including  ${\bb R}=\{x^=\}$.

\bigskip
\centerline{\bf 2. D=11 on-shell superfields}

The description of linearized 11D supergravity multiplet by superfields in the on-shell superspace (\ref{On-shellSSP}) was proposed in \cite{Galperin:1992pz} (and can be reproduced when quantizing the massless 11D superparticle \cite{Bandos:In-prep}).
It was given in terms  of a bosonic  antisymmetric tensor superfield $\Phi^{IJK}=
\Phi^{[IJK]}(x^=, \theta^-_{{q}}, v_{\alpha {q}}{}^-)$ which obeys
\begin{eqnarray}\label{D+Phi=gPsi}
D^+_{{q}}\Phi^{IJK}  = 3i\gamma^{[IJ}_{qp}  \Psi^{K]}_{p}\, , \qquad \gamma^I_{qp}\Psi^I_{p}=0\; . \qquad
\end{eqnarray}
Here $I,J,K=1,...,9$, $q,p=1,...,16$, $\gamma^I_{qp}=\gamma^I_{pq}$ are d=9 Dirac matrices, $\gamma^I\gamma^J+\gamma^J\gamma^I=\delta^{IJ}{\bb I}_{16\times 16}$, and
\begin{eqnarray}\label{D+dq:=}
D^+_{{q}}={\partial}^+_{{q}} + 2i  \theta^-_{{q}}\partial_{=} \equiv {\partial\over \partial \theta^-_{{q}}} + 2i  \theta^-_{{q}} {\partial\over \partial x^=} \;  \quad
\end{eqnarray}
is the fermionic covariant derivative obeying the d=1, ${\cal N}=16$ supersymmetry algebra $\{ D^+_{{q}},  D^+_{{p}}\} = 4i \delta_{qp}\partial_=$.

The consistency of Eq. (\ref{D+Phi=gPsi}) requires that fermionic superfield  $\Psi^{I}_{q}$ satisfies, besides
$\gamma^I_{qp}\Psi^I_{p}=0$,
\begin{eqnarray}\label{D+Psi=11D}
D^+_{{q}}\Psi^{I}_{p}  = \frac 1 {18} \left(\gamma^{IJKL}_{qp}+ 6 \delta ^{I[J}\gamma^{KL]}_{qp}  \right) \partial_{=}\Phi^{JKL}   + \quad \nonumber \\ + 2  \partial_{=}H_{IJ}\gamma^{J}_{qp}  \; , \qquad
\end{eqnarray}
with symmetric traceless $SO(9)$ tensor superfield $H_{IJ}=H_{((IJ))}$,   obeying
\begin{eqnarray}
\label{D+h=11D} D^+_{{q}}H_{IJ} = i \gamma^{(I}_{qp} \Psi^{J)}_{p} , \qquad H_{IJ}= H_{JI}, \quad H_{II}=0 \; . \;
\end{eqnarray}
The leading component of this bosonic  superfield,
 $h_{IJ}(x^=, v_{\alpha q}^{\; -})= H_{IJ}\vert_{\theta^-_q=0}$,    describes the on-shell degrees of freedom of the 11D graviton (see \cite{Galperin:1992pz} for more details).

One can collect all the above on-shell superfields in
\begin{eqnarray}\label{PsiQ=11D}
\Psi_Q (x^= ,  v_{\alpha q}^{\; -}; \theta^-_q) =\left\{ \Psi_{ Iq}\, , \Phi_{[IJK]}\, , \, H_{((IJ))}\,\right\}, \qquad
\end{eqnarray}
with multiindex $Q$ taking 128(=144-16) 'fermionic' and 128=84+44 'bosonic values', $Q= \left\{ Iq\, , [IJK]\, , \, ((IJ))\,\right\}$.
The set of equations (\ref{D+Psi=11D}), (\ref{D+Phi=gPsi}) and (\ref{D+h=11D}) can be unified in
\begin{eqnarray}\label{D+PsiQ=11D}
D^+_q \Psi_Q = \Delta_{Q\, qP} \Psi_P,  \qquad
\end{eqnarray}
where the operator $ \Delta_{Q\, qP}$ can be easily read off Eqs. (\ref{D+Psi=11D}), (\ref{D+Phi=gPsi}) and (\ref{D+h=11D}). It  contains differential operator $\partial_=$ when $Q=Iq$ and is purely algebraic otherwise. This difference is diminished when passing to the Fourier images of the superfields
with respect to $x^=$ coordinate, $ \Psi_Q (\rho^\# ,  v_{\alpha q}^{\; -}; \theta^-_q)= \frac 1 {2\pi} \int dx^= \, \exp( i \rho^\# x^= ) \,  \Psi_Q (x^= ,  v_{\alpha q}^{\; -}; \theta^-_q)$. These obey the same equation (\ref{D+PsiQ=11D}) but with $\partial_= \mapsto -i \rho^\# $
and
\begin{eqnarray}\label{D+q=p}
D^+_{{q}}={\partial}^+_{{q}} + 2 \rho^\# \theta^-_{{q}}\;  . \quad
\end{eqnarray}

As we have already noticed, the set of Eqs. (\ref{D+Psi=11D}), (\ref{D+Phi=gPsi}) and (\ref{D+h=11D}), collected in (\ref{D+PsiQ=11D}), are dependent. We can choose any of them and reproduce two others  from its consistency conditions. Passing to Fourier image makes natural to choose the fermionic superfield as fundamental and to describe the linearized 11D supergravity by the equation
\begin{eqnarray}\label{D+Psi=11Dp}
& D^+_{{q}}\Psi^{I}_{p}  = - \frac {i\rho^{\#}} {18} \left(\gamma^{IJKL}+ 6 \delta ^{I[J}\gamma^{KL]}  \right){}_{qp} \Phi^{JKL}   -\quad \nonumber \\ & - 2i\rho^{\#}  H_{IJ}\gamma^{J}_{qp}  \; . \qquad
\end{eqnarray}

Eqs. (\ref{D+PsiQ=11D}) (i.e. the set of Eqs. (\ref{D+Phi=gPsi}), (\ref{D+Psi=11D}) and (\ref{D+h=11D})) and  $\gamma^I_{{qp}}\Psi^{I}_{p}=0$ play the role of D=4 helicity constraint (\ref{UPhi=1}). Then it is natural to expect that an on-shell tree superfield amplitude should satisfy essentially the same set of equations  for each of the scattered particles.


\bigskip

\centerline{\bf 3. Tree on-shell amplitudes in D=11}




The tree on-shell $n$-particle scattering amplitudes can be described as a function in a direct product of  $n$ copies of the on-shell superspace (\ref{On-shellSSP})
 \begin{eqnarray}\label{Sf-Amp-Q}
{\cal A}^{(n)}_{Q_1... Q_n}( k_1, \theta^-_{1}; ...; k_n, \theta^-_{n}) \equiv  {\cal A}^{(n)}_{... Q_l...}( ...; k_l, \theta^-_{l}; ...) \equiv \nonumber \\ \equiv
{\cal A}^{(n)}_{ ... Q_l... }(...; \rho^{\#}_{(l)}; v^-_{{q}(l)} ; \theta^-_{{q}(l)}; ...  )\; , \qquad
\end{eqnarray}
carrying $n$ multi-indices $Q_l= \left\{ I_lq_l\, ,  [I_lJ_lK_l]\, , \, ((I_lJ_l))\, \right\}$ (see (\ref{PsiQ=11D})).
As indicated in (\ref{Sf-Amp-Q}), for shortness  we often write the bosonic argument of the amplitude as
$k_{(l)}^a$ instead of $\rho^{\#}_{(l)}; v^-_{{q}(l)} $ (implying that $k_{(l)}^a$ is expressed in terms of these by (\ref{k=pv-v-11}), where $\rho^{\#}_{(l)}$ is allowed to be negative). We will also omit the arguments of the amplitude  when this does not produce a confusion.

The set of equations for the 11D amplitudes, playing the role of D=4 helicity constraints (\ref{UcAn=cAn}), includes, besides  the $\gamma$-tracelessness on every 'fermionic' multiindex $I_lq_l$, \begin{eqnarray}\label{gIcA=0}
\gamma^{I_l}_{p_lq_l}{\cal A}_{\ldots I_{(l)}q_{(l)}\ldots }=0,  \qquad
\end{eqnarray} the equation
\begin{eqnarray}\label{D+cAQ=11D}
D^+_{q(l)} {\cal A}_{\ldots Q_{(l)}\ldots }= (-)^{\Sigma_l} \Delta_{Q_l\, q  P_{(l)}} {\cal A}_{\ldots P_{(l)}\ldots }& ,  \qquad
\end{eqnarray}
where $ \Delta_{Q_l\, q  P_{(l)}}$ is the same as in (\ref{D+PsiQ=11D}) ({\it i.e.} can  be read off
(\ref{D+Psi=11Dp}), (\ref{D+Phi=gPsi}) and (\ref{D+h=11D})), but acting on variables and indices corresponding to $l$-th particle,  and ${\Sigma_l}$
can be  defined as
the number of fermionic, $I_jq_j$, indices among  $Q_1, \ldots Q_{(l-1)}$,
{\it i.e.}
\begin{eqnarray}\label{Sl:=}
& {\Sigma_l} =\sum\limits_{j=1}^{l-1}  \frac {(1-(-)^{\varepsilon ({Q_j})})}{2}\; ,  \quad \left\{
{}^{\varepsilon ({[I_jJ_jK_j]})=0=\varepsilon ( \,{((I_jJ_j))}\, )\; ,}
_{\varepsilon ({I_jq_j})=1\; .} \right. \qquad
\end{eqnarray}
In particular, when $Q_l= I_lp_l$,  Eq. (\ref{D+cAQ=11D}) reads
\begin{widetext}
\begin{eqnarray}\label{Df-AIq}
 & (-)^{\Sigma_l} D^{+(l)}_{{q}_l}  {\cal A}^{(n)}_{Q_1... \, I_l p_l\, ... Q_n}
 =   - 2i  \rho^\#_{(l)} \gamma_{J_l\, qp}{\cal A}^{(n)}_{Q_1... ((I_lJ_l))... Q_n}  -\frac {i} {18} \rho^\#_{(l)}  \left(\gamma^{I_lJ_lK_lL_l}_{qp}+ 6 \delta ^{I_l[J_l}\gamma^{K_lL_l]}_{qp}  \right)
 {\cal A}^{(n)}_{Q_1... [J_lK_lL_l]... Q_n }\; . \qquad
\end{eqnarray}
\end{widetext}


\centerline{\bf 4. Generalized BCFW deformation in D=11 }


To write the generalized BCFW recurrent relations in D=11 we have to define the  generalized BCFW deformation of bosonic and fermionic variables of the above described  11D on-shell superfield formalism.

As in the original 4D construction \cite{Britto:2005fq}, the deformation  of say the $1$-st and the $n$-th particle variables should imply  the opposite shift of their light-like momenta
\begin{eqnarray}\label{BCWF=hp}
\widehat{k_{(1)}^{a}}= k_{(1)}^{a} -z q^a \; , \qquad \widehat{k_{(n)}^{a}}= k_{(n)}^{a} +z q^a \; , \qquad
\end{eqnarray}
on a light-like vector $q^a$  orthogonal to both $ k_{(1)}^{a}$ and $ k_{(n)}^{a}$,
\begin{eqnarray}\label{qq=0=}
& q_aq^a=0\; , \qquad q_ak_{(1)}^{a} =0\; , \qquad q_ak_{(n)}^{a}= 0\; , \qquad
\end{eqnarray}
multiplied by an arbitrary complex number  $z\in {\bb C}$ \cite{Britto:2005fq} (10D construction of \cite{CaronHuot:2010rj} used real  $z\in {\bb R}$). Eqs. (\ref{qq=0=}) guarantee that the deformed momenta remain light-like
\begin{eqnarray}\label{BCWF=hp}
& ({k_{(1)}})^{2}= 0=({k_{(n)}})^{2} \quad \Rightarrow \quad  (\widehat{k_{(1)}})^{2}= 0= (\widehat{k_{(n)}})^{2}.  \quad
\end{eqnarray}
Thus  the amplitude  depending on these, instead of original  $ k_{(1)}^{a}$ and $ k_{(n)}^{a}$,
$\; {\cal A}_{z\, Q_1...Q_n}(\widehat{k_{(1)}} \,, \theta^-_{(1)}; \ldots \widehat{k_{(n)}}, \theta^-_{(n)})$, remains an on-shell amplitude.

In D=4 the deformation of the momenta (\ref{BCWF=hp}) results from the following deformation of the bosonic spinors entering the Penrose representation (\ref{p=ll=4D})
 \begin{eqnarray}\label{BCFWln1=4D}
\widehat{\lambda^{A}_{(n)}} = \lambda^{A}_{(n)} + z \lambda^{A}_{(1)} , \qquad \widehat{\bar{\lambda}{}^{\dot A}_{(1)}} = \bar{\lambda}{}^{\dot A}_{(1)}- z \bar{\lambda}{}^{\dot A}_{(n)} , \qquad  \end{eqnarray}
In D=11 (\ref{BCWF=hp})
results from the following deformation of  the associated spinor moving frame variables
\begin{eqnarray}\label{BCWF=vnM}
\widehat{v_{\alpha {q}(n)}^{\; -}}= v_{\alpha {q}(n)}^{\; -} + z \;
v_{\alpha {p}(1)}^{\; -} \; {\bb M}_{ {p}{q}} \;
\sqrt{{\rho^{\#}_{(1)}}/{\rho^{\#}_{(n)}}}
\; , \qquad \\ \label{BCWF=v1M} \widehat{v_{\alpha {q}(1)}^{\; -}}= v_{\alpha {q}(1)}^{\; -} - z \; {\bb M}_{{q}{p}}\;  v_{\alpha {p}(n)}^{\; -}\;
 \sqrt{ {\rho^{\#}_{(n)}}/{\rho^{\#}_{(1)}}}
\;  \qquad \end{eqnarray}
which enter the Penrose-like constraints (\ref{k=pv-v-11}),
\begin{eqnarray}\label{ki=pv-v-11}
k^a_{(i)}
\Gamma_{a\alpha\beta}= 2\rho^{\#}_{(i)} v_{\alpha q(i)}^{\; -} v_{\beta q(i)}^{\; -}, \quad \nonumber \\ {}
 k_{a(i)} \delta_{{q}{p}}= \rho^{\#}_{(i)} v^-_{{q}(i)} \tilde{\Gamma}_{a}v^-_{{p}(i)}\; . \qquad
\end{eqnarray}
The energy variables ${\rho^{\#}_{(i)}}$ are not deformed.
The matrix $ {\bb M}_{{q}{p}}$ is constructed from the light-like vector $q^a$ of (\ref{BCWF=hp})
\begin{eqnarray}\label{Mpq=}
{\bb M}_{{q}{p}} =-   {q^a}\,  {(v_{{q}(1)}^{\; -} \, \tilde{\Gamma}_av_{{p}(n)}^{\; -})} \, {\sqrt{\rho^{\#}_{(1)}\rho^{\#}_{(n)} }} /{ (k_{(1)}k_{(n)}) } \;  \qquad
\end{eqnarray}
({\it cf.} with 10D relations in \cite{CaronHuot:2010rj}), with ${16}  k_{(i)}^a= {\rho^{\#}_{(i)}} v_{{q}(i)}^{\; -}\tilde{\Gamma}^a v_{{q}(i)}^{\; -}$ (see (\ref{ki=pv-v-11})),
and is nilpotent
\begin{eqnarray}\label{MMT=0}
{\bb M}_{rp} {\bb M}_{r{q}} =0\;  ,\qquad {\bb M}_{{q}r } {\bb M}_{{p}r} =0\;  ,\qquad
\end{eqnarray}
due to (\ref{qq=0=}).
This nilpotent matrix enters also the deformation rules of the fermionic coordinates
\begin{eqnarray}\label{BCFW=thn}
\widehat{ \theta^-_{{p}(n)}}= \theta^-_{{p}(n)}+ z \,\theta^-_{{q}(1)} \, {\bb M}_{{q}{p}} \, \sqrt{ {\rho^{\#}_{(1)}}/{\rho^{\#}_{(n)}}} \; , \qquad \\ \label{BCFW=th1}
\widehat{ \theta^-_{{q}(1)}}= \theta^-_{{q}(1)}- z  \, {\bb M}_{{q}{p}} \, \theta^-_{{p}(n)}\, \sqrt{ {\rho^{\#}_{(n)}}/{\rho^{\#}_{(1)}}}  \; . \qquad \end{eqnarray}
These can be also written as
\begin{eqnarray}\label{BCFW=thl}
\widehat{ \theta^-_{{p}(i)}} &=& e^{-zD^+_{(1)}{\bb M}\theta^-_{(n)} -z \theta^-_{(1)}{\bb M}D^+_{(n)} }\; \theta^-_{{p}(i)}\; , \qquad
 \end{eqnarray}
 where the covariant fermionic derivatives $D^+_{q(i)}$ are defined in (\ref{D+q=p}).
 Their deformation
\begin{eqnarray}
\label{BCFW=D+l}
&& \widehat{ D^+_{{q}(i)}} = \\ \nonumber  && e^{-zD_{(1)}{\bb M}\theta_{(n)} -z \theta_{(1)}{\bb M}D_{(n)} }\; D^+_{{q}(i)} e^{zD_{(1)}{\bb M}\theta_{(n)} +z \theta_{(1)}{\bb M}D_{(n)} }\;   \end{eqnarray}
is similar to the deformation of 8d Clifford algebra valued variables in the 10D construction of \cite{CaronHuot:2010rj}.

\bigskip

\centerline{\bf 5. Generalized BCFW recurrent relations}
\centerline{\bf  for tree amplitudes in $D=11$}

The deformed tree amplitude is defined as an amplitude depending on deformed momenta and fermionic coordinates. We denote it by
\begin{eqnarray}\label{cAz:=def}
\widehat{{\cal A}_z^{(n)}}{}_{_{... {Q_l} ...}}:= {{\cal A}}^{(n)}_z{}{_{Q_1... {Q_l} ...Q_n}}( \widehat{k_{(1)}}, ...; \widehat{k_{(l)}}, \widehat{\theta^-_{(l)}}; ..., \widehat{\theta^-_{(n)}}) \quad \\ \nonumber
= {{\cal A}}_z^{(n)}{}_{ Q_1... Q_n}(\widehat{k_{(1)}}, \widehat{\theta^-_{(1)}};k_{(2)} , \ldots ,   \theta^-_{(n-1)}; \widehat{k_{(n)}}, \widehat{\theta^-_{(n)}})\; ,
\end{eqnarray}
where in the last line it is assumed that the deformed momenta correspond to $1$-st and $n$-th of the scattered  particles (so that  $\widehat{k_{(l)}}, \widehat{\theta^-_{(l)}}= {k_{(l)}}, {\theta^-_{(l)}}$ for $l=2,...,(n-1)$), and the subscript $z$ indicates the parameter used in
this deformation
(\ref{BCWF=vnM})--(\ref{BCFW=th1}).
Notice that deformed amplitudes (\ref{cAz:=def}) satisfy, besides the gamma-tracelessness  (\ref{gIcA=0}), Eqs. (\ref{D+cAQ=11D}) with deformed derivatives (\ref{BCFW=D+l}),
\begin{eqnarray}\label{D+chAQ=11D}
\widehat{D^+_{q(l)}} \widehat{{\cal A}}_{z\, Q_1 \ldots Q_{(l)}\ldots }= (-)^{\Sigma_l} \Delta_{Q_l\, q  P_{(l)}}
\widehat{{\cal A}}_{z, Q_1\ldots P_{(l)}\ldots }& .  \qquad
\end{eqnarray}
In particular,
\begin{eqnarray}\label{Df-hcAIJK}
(-)^{\Sigma_l} \widehat{D^+_{q_l(l)}}  \widehat{{\cal A}}_{z\, ... [I_lJ_lK_l]... }=
 \, 3i \gamma_{[J_lK_l| q_lp_l} \, \widehat{{\cal A}}_{z\,... \, |I_l] p_l\, ... } , \quad \\
 (-)^{\Sigma_l} \widehat{D^+_{q_l(l)}}  \widehat{{\cal A}}_{z\, ... ((I_lJ_l))... }=
 \, i \gamma_{ q_lp_l((I_l|} \, \widehat{{\cal A}}_{z\,... \, |J_l)) p_l\, ... } . \qquad
\end{eqnarray}
The proposed BCFW-type recurrent relation for tree superfield amplitudes of 11D supergravity reads
\begin{widetext}
\begin{eqnarray}\label{cA-Sf=rBCFW11D}
&{\cal A}^{(n)}_{Q_1\ldots Q_n}& ({k_1},\theta^-_{(1)};k_2 , \theta^-_{(2)};\ldots ;{k_n},\theta^-_{(n)})= \qquad\nonumber \\ & =&  \sum\limits_{l}^n \frac{(-)^{\Sigma_{(l+1)}}}{64 (\widehat{\rho}{}^{\#}(z_l))^2}D^+_{{q}(z_l)}\left(
\widehat{{\cal A}}{}^{(l+1)}_{z_l \; Q_1\ldots Q_l \; Jp}(\widehat{k_1}, \widehat{\theta^-_{(1)}};k_2, \theta^-_{(2)}; \ldots ;{k_l}, \theta^-_{(l)};\widehat{P_{l}}(z_l),\Theta^-) \right. \times
\qquad \\ \nonumber
&&  \left. \qquad
\times \frac {1}{(P_{l})^2} \overleftrightarrow{D^+}_{{q}(z_l)}  \widehat{{\cal A}}{}^{(n-l+1)}_{z_l\; Jp\;  Q_{l+1}\ldots Q_n}(-\widehat{P_{l}}(z_l),\Theta^-; k_{l+1}, \theta^-_{(l+1)};\ldots ; k_{n-1}, \theta^-_{(n-1)}; \widehat{k_n}, \widehat{\theta^-_{(n)}})\right)\vert_{\Theta^-=0}
\; .
\end{eqnarray}
 \end{widetext}
Here
\begin{eqnarray}\label{kSl=} {P_{l}^a}=- \sum\limits_{m=1}^l  {k_m^a}\; , \qquad
 \\
 \label{hkSl=}
 \widehat{P_{l}^a}(z)=- \sum\limits_{m=1}^l  \widehat{k_m^a}(z)
= {P_{l}^a} - z  q^a\; ,  \qquad
\\
 \label{zl:=}
z_l:= {P_{l}^a P_{l \,a}}/ ({2P_{l}^b q_b})\; , \qquad
\end{eqnarray}
with $q^a$ obeying   (\ref{qq=0=}) and (\ref{Mpq=})\footnote{ One can check that (\ref{Mpq=}) can be formally resolved by
$q^a= \sqrt{\rho^{\#}_1\rho^{\#}_n}  v_{q(1)}^{-}\tilde{\Gamma}^a{\bb M}_{{q}{p}}v_{{p}(n)}^{-}/8\;$.}.
 Eq. (\ref{hkSl=}) implies that
$(\widehat{P_{l}}(z))^2= (P_{l})^2 -2z P_{l}\cdot q$, so that
$\widehat{P_{l}^a}(z_l)$ is light--like
\begin{eqnarray}\label{hkSl2zl=0}
(\widehat{P_{l}}(z_l))^2=0\; , \qquad z_l:= {(P_{l})^2}/ ({2P_{l}\cdot q})\; . \;
\end{eqnarray}
As a result, firstly, both amplitudes in the {\it r.h.s.} of (\ref{cA-Sf=rBCFW11D}) are on the mass shell, and secondly we can express $\widehat{P_{l}^a}(z)$ in terms of assiciated spinor movig frame variables  $v_{\alpha {q}}^{\; -}(z_l):= v_{\alpha {q}}^{\; -}{}_{\hat{P_l}(z_l)}$ and  energy $\pm\widehat{\rho}{}^{\#}(z_l)$
(see  (\ref{k=pv-v-11}))
\begin{eqnarray}\label{hkSlzl=}
\widehat{P_{l}{}^a}(z_l)\Gamma_{a\alpha\beta} = 2 \widehat{\rho}{}^{\#}(z_l)\,  v_{\alpha {q}}{}^{\!\! -}(z_l) v_{\beta {q}}{}^{\!\! -}(z_l) \; , \qquad \nonumber \\
\widehat{P_{l}}{}^a(z_l)\delta_{qp}  = \widehat{\rho}{}^{\#}(z_l)\,  v^{-}_{{q}} (z_l)\tilde{\Gamma}^{a} v^{-}_{p} (z_l)\; . \qquad
\end{eqnarray}
This $\widehat{\rho}{}^{\#}(z_l)$ enters the denominator of the terms in r.h.s. of (\ref{cA-Sf=rBCFW11D}) (which is needed to simplify the relation between amplitude and superamplitude).

Actually, the bosonic arguments of the on-shell amplitudes are energies $\rho^\#_{(i)}$ and
$v_{\alpha (i)}^{\; -}$
 related to light-like momenta $k_{(i)}^a$  by (\ref{ki=pv-v-11}), and the above
   $v_{\alpha {q}}^{\; -}(z_l)$ and  $\pm\widehat{\rho}{}^{\#}(z_l)$;
  just for shortness in (\ref{cA-Sf=rBCFW11D}), following (\ref{Sf-Amp-Q}), we hide this writing instead the  dependence on the momenta.

Finally, $D^+_{{q}(z_l)}$ in (\ref{cA-Sf=rBCFW11D}) is the covariant derivative with respect to $\Theta^-_{{q}}$
constructed with the use of $\widehat{\rho}{}^{\#}(z_l)$ of (\ref{hkSlzl=}),
\begin{eqnarray}\label{D+=Sl}
D^+_{{q}(z_l)}= \frac {\partial }  {\partial \Theta^-_{{q}}} + 2\widehat{\rho}{}^{\#}(z_l)  \Theta^-_{{q}}
\; . \qquad
\end{eqnarray}

Notice that the structure of the
r.h.s. of  (\ref{cA-Sf=rBCFW11D}),
\begin{eqnarray}\label{D+=Sl}
& D^+_{{q}}\left(
{\cal A}_{... Jp}\overleftrightarrow{D^+}_{{q}} {\cal A}_{ Jp\;  \ldots }\right)\vert_{\Theta^-=0}\equiv
\\ \nonumber & \equiv  D^+_{{q}}\left(
{\cal A}_{\ldots  Jp}D^+_{{q}} {\cal A}_{Jp\; \ldots } - (-)^{\Sigma_l}
D^+_{{q}} {\cal A}_{\ldots  Jp}\;  {\cal A}_{Jp\;  \ldots }\right)\vert_{0},
\end{eqnarray}
 can be treated as an integration over the fermionic  variable $\Theta^-_q$ in (\ref{D+=Sl}) with an exotic measure similar to one used in \cite{Tonin:1991ii,Tonin:1991ia} to construct a worldsheet superfield formulation of the heterotic string (see  \cite{Zupnik:1989bw} for formal discussion on superspace measures).

To argue that there is no contribution to the r.h.s. of  (\ref{cA-Sf=rBCFW11D}) of a pole at $|z|\mapsto \infty$ , we can use the line of arguments presented in \cite{CaronHuot:2010rj} for 10D case, which refers on the case when external momenta lays in some 4d subspace of spacetime and on the original proof
of \cite{Britto:2005fq} which was extended to ${\cal N}=8$ supergravity in \cite{Brandhuber:2008pf,ArkaniHamed:2008gz,Heslop:2016plj}.

The calculation of sample  tree superamplitudes of 11D supergravity with the use of the above BCFW-type recurrent relations (\ref{hkSlzl=}), and  generalization of these to  loop amplitudes will be the subject of subsequent work.  See  supplemental material to this paper \footnote{Supplemental material can be found at 
  http://link.aps.org/supplemental/10.1103/PhysRevLett. 118.031601.  The references \cite{Deser:2000xz} --- \cite{Sannan:1986tz} are cited their. } for some technicalities needed to proceed with explicit superamplitude calculations. 


{\bf Acknowledgements}
This work has been supported in part by the Spanish MINECO grant FPA2012-35043-C02-01,  partially financed
with FEDER/ERDF (European Regional Development Fund of the European Union), by the Basque Government Grant IT-979-16, and the Basque Country University program UFI 11/55.
The author is thankful to Theoretical Department of CERN for hospitality and support of his  visits  at different stages of this project, and  to Luis Alvarez-Gaume, Boris Pioline, Emeri Sokatchev  and Paolo Di Vecchia for useful discussion on related topics during these visits.


\end{document}